\documentclass[11pt,a4paper]{article}
\usepackage{jheppub}
\usepackage{cases}
\usepackage{graphicx}
\usepackage{dcolumn,color}
\usepackage{bm}
\usepackage{epstopdf}
\usepackage{epsfig,subfigure}
\usepackage{color,framed}
\usepackage{amsmath,amsthm,amssymb} 

\usepackage[table]{xcolor}

\title{\boldmath Thick brane in Rastall gravity }

\author[a,b]{Yi Zhong,}
\author[c]{Ke Yang,}
\author[b,1]{and Yu-Xiao Liu\note{Corresponding author.}}

\affiliation[a]{ School of Physics and Electronics Science,\\
Hunan Provincial Key Laboratory of High-Energy Scale Physics and Applications,\\
             Hunan University, Changsha 410082, P. R. China}
\affiliation[b]{Lanzhou Center for Theoretical Physics,
Key Laboratory of Theoretical Physics of Gansu Province,
School of Physical Science and Technology,
Lanzhou University, Lanzhou 730000, China}
\affiliation[c]{School of Physical Science and Technology, Southwest University, Chongqing 400715, China}

\emailAdd{zhongy@hnu.edu.cn}
\emailAdd{keyang@swu.edu.cn}
\emailAdd{liuyx@lzu.edu.cn}

\abstract{In this work, thick branes in Rastall gravity are investigated. Three types of  maximally symmetric thick brane models are constructed and the linear tensor perturbation is analyzed. In the flat brane model, the tensor modes of the perturbation are either unstable or nonlocalizable for a nonvanishing Rastall parameter.  In the de Sitter brane model, only the
ground state of the tensor mode is localized. In the anti-de Sitter brane model, the number
of the bound tensor states is infinity.  For both the de Sitter and anti-de Sitter brane models, the
condition of stability for the Rastall parameter is obtained.}

\begin{document}
\maketitle

\section{Introduction}

It has been an attractive idea for the last two decades that our spacetime may have extra dimensions and the observable universe could be a brane in a higher-dimensional spacetime. In the braneworld scenario, it is possible to explain the hierarchy between the Planck scale and the electroweak scale \cite{ArkaniHamed:1998rs,Randall:1999ee,Randall:1999vf}. The braneworlds in these models are  geometrically thin. However, in a viable model, a braneworld should have a thickness.
Inspired by the domain wall in Ref. \cite{Rubakov:1983bb}. The thick braneworld generated by a background scalar field was proposed \cite{Gremm:1999pj,Csaki:2000fc,PhysRevD.62.044017,Kobayashi:2001jd,PhysRevD.66.024024,Bazeia:2002sd,Andrianov:2005hm,Afonso:2006gi,Neupane:2010ey}.
Furthermore, since general relativity suffers from both  phenomenological and theoretical problems such as dark energy problem and singularity problem, various modified theories of gravity have been proposed, and braneworlds in various modified gravities have been investigated \cite{Liu:2017gcn}.
It was found that the branes may have inner structure in some theories such as $f(T)$ gravity \cite{Yang:2012hu} and  mimetic gravity \cite{Zhong:2017uhn}. Pure geometric thick branes were found in modified gravities \cite{Arias:2002ew,Barbosa-Cendejas:2005vog,Barbosa-Cendejas:2006cic,Zhong:2015pta} as well. See Refs. \cite{Giovannini:2001ta,Cho:2001nf,Heydari-Fard:2007dos,Nozari:2008ny,Parry:2005eb,Afonso:2007gc,Dzhunushaliev:2009dt,Koivisto:2005yk,Bazeia:2013oha,Chakraborty:2015taq,Liu:2012rc,Yang:2017puy,Bogdanos:2006qw,Farakos:2007ua,Fiorini:2013hva} for  more recent works of braneworld in modified gravities and Ref. \cite{Liu:2017gcn} for a review.

In general relativity and many modified gravities, matter fields are minimally coupled to gravity, and the energy-momentum tensor conserves as a result of the Noether theorem. However, the energy-momentum conservation law is violated in the particle creation process in cosmology. Accordingly, it is reasonable to relax the conservation condition of the energy-momentum tensor.
 P. Rastall proposed such kind of modified gravity \cite{PhysRevD.6.3357}. In this theory, the covariant divergence of the energy-momentum tensor is assumed to be proportional to the scalar curvature, i.e. $\nabla^{\mu}T_{\mu\nu}=\lambda \nabla_{\nu}R$. Therefore, the energy-momentum conservation law is recovered in a flat spacetime.
  This theory has attracted much attention recently, and cosmology and compact stars have been investigated  in Rastall gravity \cite{BezerradeMello:2014okn,Heydarzade:2017wxu,Darabi:2017tay,Xu:2017vse,Darabi:2017coc,Das:2018dzp,Tang:2019dsk,Li:2019jkv,Khyllep:2019odd,Ghosh:2021byh,Haghani:2022lsk,Shahidi:2021lxt}. Constrains of the Rastall parameter $\lambda$ from the galaxy-scale strong gravitational lensing and rotation curves of low surface brightness galaxies were given in Refs. \cite{Tang:2019dsk,Li:2019jkv}.

  It was shown that Rastall gravity and its generalization is equivalent to the $f(R,T)$ gravity with the Lagrangian given by $f(R,T)= R+\alpha T$ \cite{Shabani:2020wja}, where $R$ is the Ricci scalar and $T$ the trace of stress-energy tensor. The thick brane model in $f(R,T)$ gravity has been considered in Ref. \cite{Gu:2016nyo}, and the full linear perturbations of the model were analyzed. It was shown that only the scalar mode of the perturbation differs from the standard case in general relativity.
In this work, we will consider the thick branes in Rastall gravity and investigate the localization of gravity and the stability against tensor perturbation.  Three types of
maximally symmetric thick branes will be considered. We will show that the thick braneworlds in Rastall gravity are quite different from that in $f(R,T)$ gravity in the context of tensor perturbation.

This work is organized as follows. In Sec. 2, we solve the thick brane generated by a scalar field in Rastall gravity.
In Sec. 3,  the linear perturbation of the background metric is analyzed, and its Kaluza-Klein modes are solved when $| \lambda |\ll  1$.
Finally, we give out the conclusion and discussion.



\section{The Model}
We start with a scalar filed $\phi$ of which the energy-momentum tensor $T_{MN}=\partial_M \phi \partial_N \phi-g_{MN}\left(\frac{1}{2}\partial^{M} \phi \partial_{M}
        \phi +V(\phi)\right)$ is not conserved, and the covariant divergence is related to the scalar curvature $R$ by
    \begin{eqnarray}
    	\label{covdT}
        \nabla^{M}T_{MN}=\lambda \nabla_{N}R,
    \end{eqnarray}
where the constant $\lambda$ is a measure of the tendency of the nonminimal coupling of the matter field and gravity. It is reasonable to assume that $| \lambda | \ll 1$. Throughout this paper, we take the geometrized units in which the gravitational constant $\kappa^2=1$,  and the indices $M,N\cdots=0,1,2,3,5$ denote the coordinates in the bulk, and $\mu,\nu\cdots=0,1,2,3$ denote the ones on the brane.
The field equation of the generalized Rastall gravity in $5$-dimensional spacetime is
    \begin{eqnarray}
        \label{Rastal eq}
        G_{MN}+\lambda R g_{MN}=T_{MN},
    \end{eqnarray}
or
        \begin{eqnarray}
        \label{Rastal eq2}
        R_{MN}=T_{MN}+\alpha T g_{MN},
    \end{eqnarray}
    where $\alpha=\frac{1-2\lambda}{10\lambda-3}$. Note that $\lambda\neq\frac{3}{10}$ as the trace of Eq. (\ref{Rastal eq}) gives $(5\lambda-\frac{3}{2})R=T$.
The metric is assumed as
    \begin{eqnarray}
        \label{metric}
       ds^2=a^2(z)(\gamma_{\mu\nu}dx^{\mu}dx^{\nu}+dz^2),
    \end{eqnarray}
where $a(z)=\text{e}^{A(z)}$ is the  warp factor, and $\gamma_{\mu\nu}$ is the metric of four-dimensional maximally symmetric spacetime, i.e
    \begin{eqnarray}
        \label{gamma metric}
       \gamma_{\mu\nu}dx^{\mu}dx^{\nu}=\begin{cases}
   \eta_{\mu\nu}dx^{\mu}dx^{\nu},&\mbox{flat brane},\\
   -dt^2+\text{e}^{2kt}(dx_1^2+dx_2^2+dx_3^2 )  ,&\mbox{de Sitter brane},\\
   \text{e}^{2kx_3}(-dt^2  + dx_1^2+dx_2^2)+dx_3^2,&\mbox{anti-de Sitter brane}.
   \end{cases}
    \end{eqnarray}

Since the brane is maximally symmetric, the scalar field can be assumed as $\phi=\phi(z)$. With the metric assumption of (\ref{metric}), Eqs. (\ref{Rastal eq}) and (\ref{covdT}) read
    \begin{eqnarray}
        \label{eom1}
        3A'^2-\phi'^2-3A''-3\epsilon k^2 &=&0, \\
        \label{eom2}
        2\text{e}^{2A}V(\phi)+ 3(3+8\lambda)(A'^2-\epsilon k^2)+(3+16\lambda)A''
                 &=&0,	\\
        \label{eom3}
      	 -24\lambda A'^3+8 \lambda  A''' +4A'\left(2 \lambda A''+\phi'^2+6\epsilon k^2\right)-\text{e}^{2A}\phi' V'(\phi )&=&0,
    \end{eqnarray}
where the primes denote the derivatives with respect to $z$, and the constant $\epsilon=0,1,-1$ for the flat, de Sitter and anti-de Sitter branes, respectively.
Note that there are only two independent equations in Eqs. (\ref{eom1})-(\ref{eom3}), we can construct the brane model by specifying the parameter $\lambda$ and assuming the warp factor $a(z)$. The potential is solved as
        \begin{eqnarray}
    	\label{potential}
 	V(\phi) =-\frac{1}{2}\text{e}^{-2A}\left[ 3(3+8\lambda)(A'^2-\epsilon k^2)+(3+16\lambda)A'' \right] .
    \end{eqnarray}
\subsection{Case I: Flat Brane}
Firstly, we consider the flat brane. Let us assume the warp factor as
    \begin{eqnarray}
    	\label{wf1}
 	a(z) =\frac{1}{\sqrt{\beta^2 z^2 +1}} .
    \end{eqnarray}
The scalar field can be solved with Eq. (\ref{eom1}):
    \begin{eqnarray}
 	\phi(z) = \sqrt{3}\text{arctan}(\beta z).
    \end{eqnarray}

 Since Eq. (\ref{eom1}) is the same as that of the brane model in general relativity, the nonconservation of the scalar field only has an effect on the potential.
The potential $V(\phi)$ is given by
    \begin{eqnarray}
 	V(\phi) = \frac{1}{4} \beta^2 \left[(56 \lambda +15) \cos \left(\frac{2 \phi }{\sqrt{3}}\right)-3 (8 \lambda +3)\right],
    \end{eqnarray}
 which is almost the same as that of the brane solution in  general relativity except for a constant term.
     \begin{figure}[!htb]
    \begin{center}
    \subfigure[The warp factor]{
        \includegraphics[width=6cm]{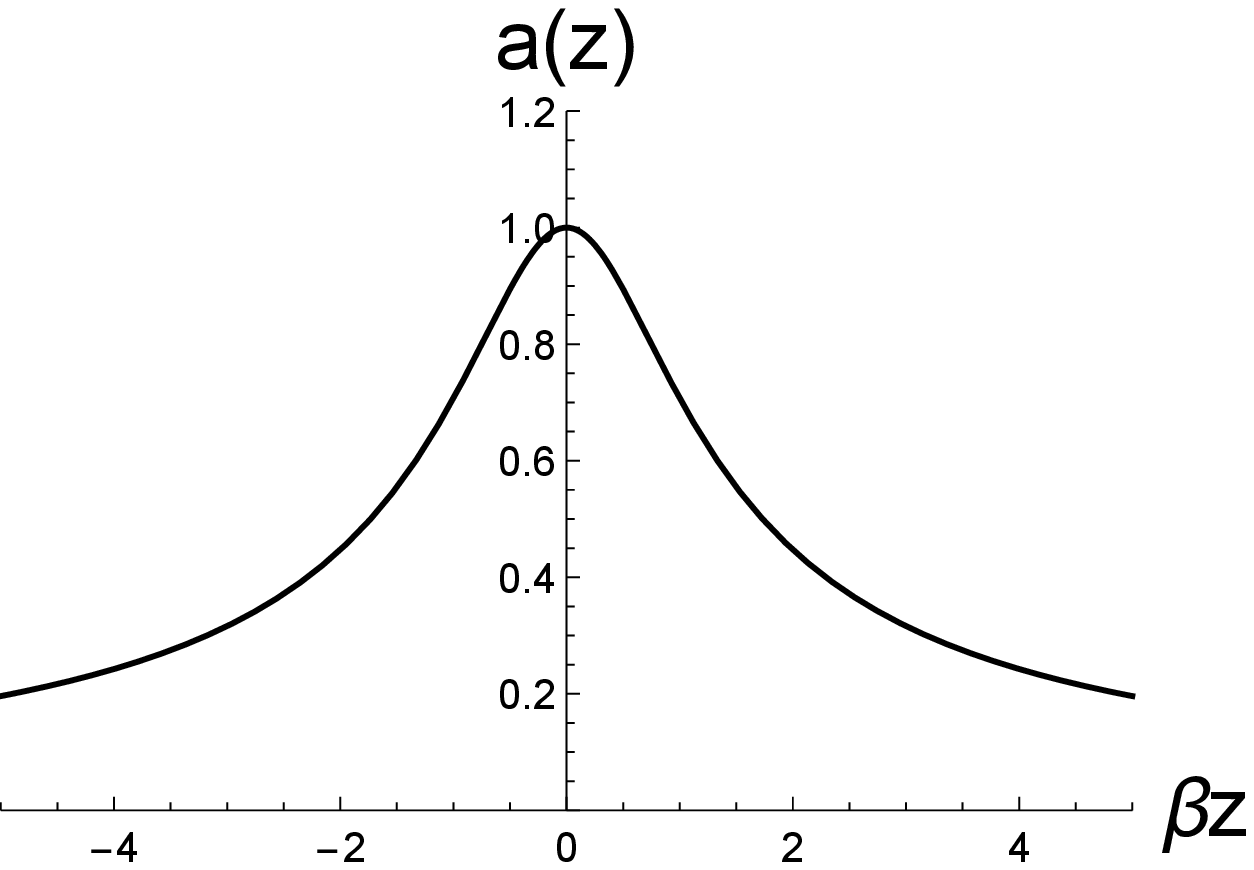}}
    \subfigure[The  scalar field]{
        \includegraphics[width=6cm]{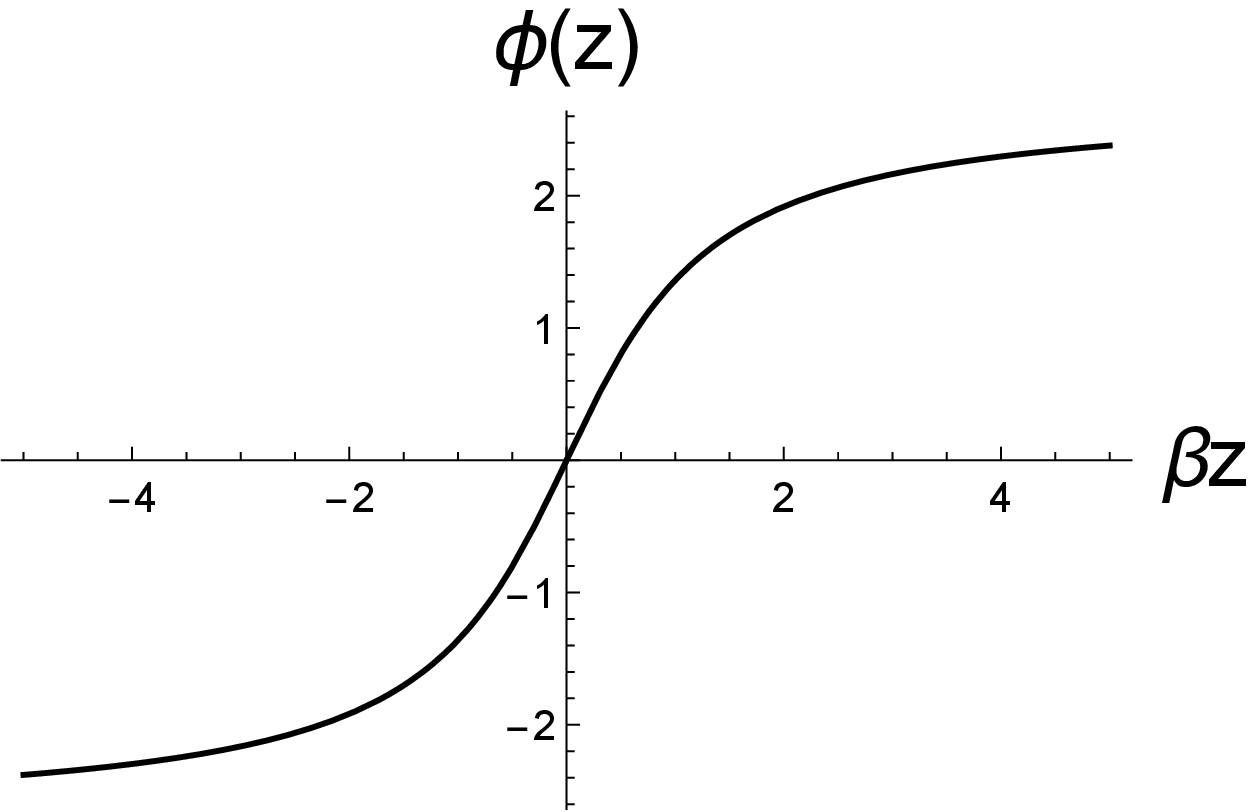}}
    \end{center}
    \caption{The shapes of the warp factor $a(z)$ and the scalar field $\phi(z)$ in the flat brane model.}\label{flat}
    \end{figure}
\subsection{Case II: de Sitter Brane}
Similarly, we assume the following warp factor of a de sitter brane,
    \begin{eqnarray}
    	\label{wf2}
 	a(z) = \text{sech}^{p}(\beta z),
    \end{eqnarray}
 with the scalar field solved as
    \begin{eqnarray}
 	\phi(z) = \phi_0 \text{arcsin}\left(\tanh(\beta z)\right),
    \end{eqnarray}
where $\phi_0=\left[3p(p-1)\right]^\frac{1}{2}$, $\beta=\frac{k}{p}$ and $0<p<1$.
The corresponding potential of the scalar field is
       \begin{eqnarray}
 	V(\phi) = \frac{1}{2} \beta ^2 p \left(16 \lambda +3 (8 \lambda +3) p+3\right) \cos ^{2 (1-p)}\left(\frac{\phi }{\phi _0}\right).
    \end{eqnarray}
    \begin{figure}[!htb]
    \begin{center}
    \subfigure[The warp factor]{
        \includegraphics[width=6cm]{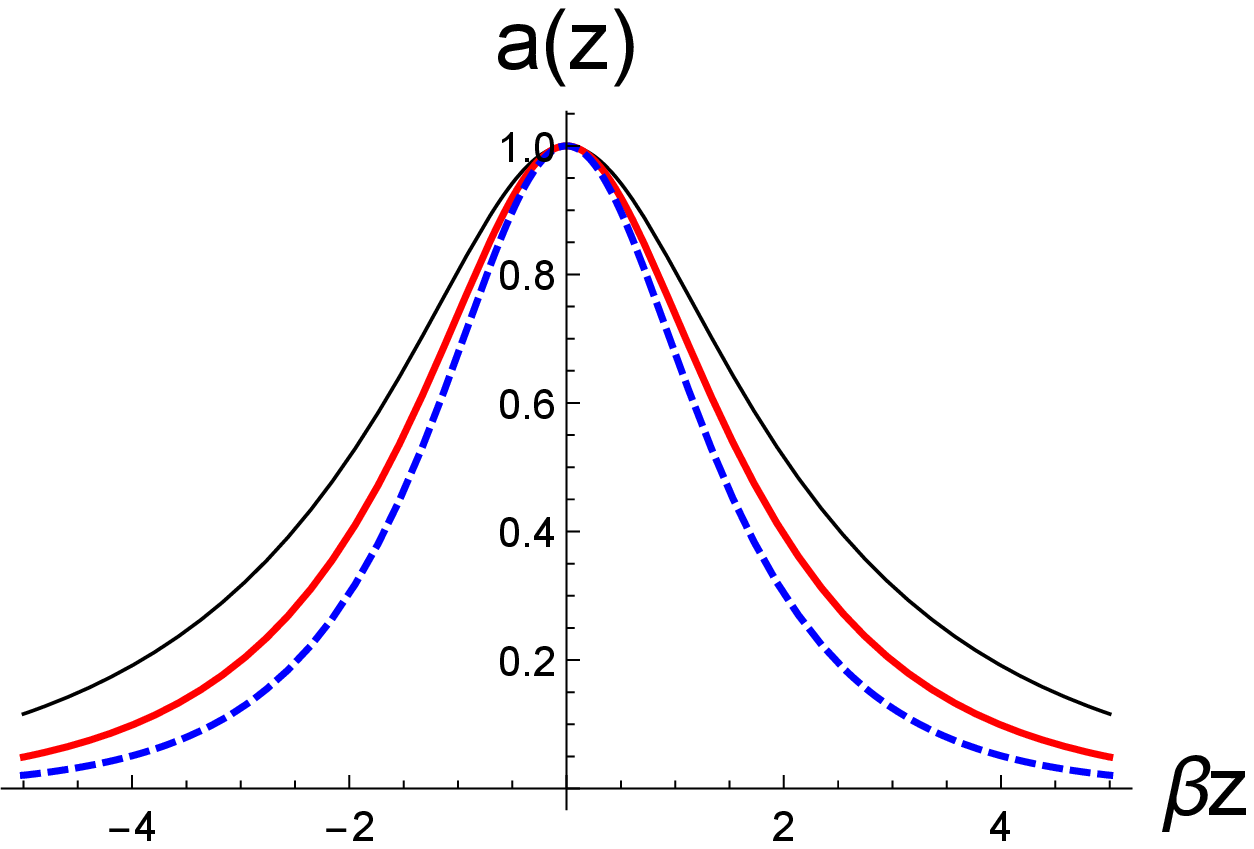}}
    \subfigure[The  scalar field]{
        \includegraphics[width=6cm]{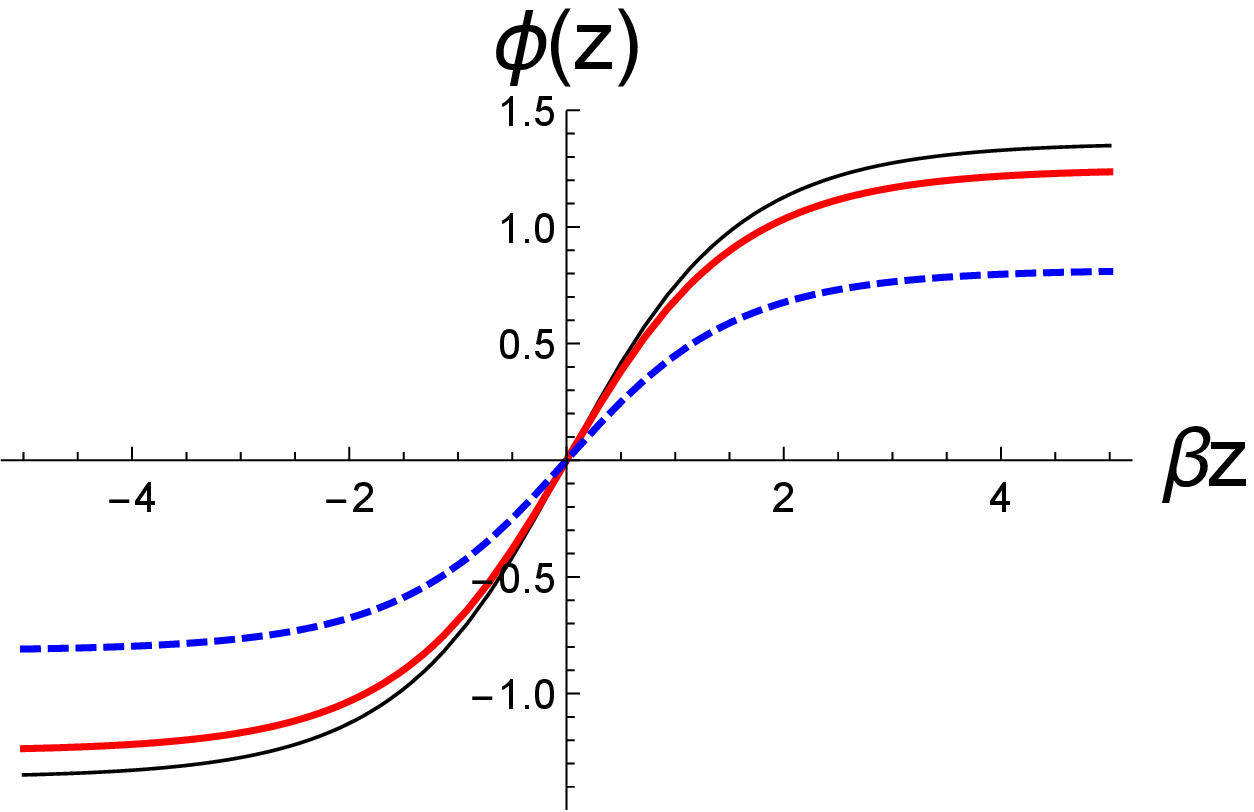}}
    \end{center}
    \caption{The shapes of the warp factor $a(z)$ and the scalar field $\phi(z)$ in the de Sitter brane model. The parameter is set as $p=0.5$ for the thin black lines, $p=0.7$ for the thick red lines, and $p=0.9$ for the dashed blue lines.}\label{flat}
    \end{figure}

\subsection{Case III: Anti-de Sitter Brane}
At last, the  warp factor of the  anti-de Sitter brane is assumed as follows \cite{PhysRevD.84.044033},
    \begin{eqnarray}
    	\label{wf3}
 	a(z) = \cos^{-p} (\beta z).
    \end{eqnarray}
The scalar field is solved as
    \begin{eqnarray}
 	\phi(z) = \phi_0 \text{arcsinh}\left(\tan(\beta z)\right),
    \end{eqnarray}
 where $\phi_0=\left[3p(1-p)\right]^\frac{1}{2}$ and $\beta=\frac{k}{p}$. The parameter $p$ satisfies $p<0$. The range of the extra dimension is  $-z_b<z<z_b$ with $z_b=\frac{p\pi}{2k}$. In this model, the metric has a naked singularity and the scalar field is  divergent at $\pm z_b$. However, it is demonstrated that these problems are acceptable  since they can be solved by lifting the bulk to
ten dimensions or by string theory \cite{PhysRevD.62.044017,PhysRevD.84.044033}.

Correspondingly,  the potential of the scalar field is given by
       \begin{eqnarray}
 	V(\phi) = -\frac{1}{2} \beta ^2 p (16 \lambda +3 (8 \lambda +3) p+3) \cosh ^{2(1-p)}\left(\frac{\phi}{\phi _0}\right).
    \end{eqnarray}

        \begin{figure}[!htb]
    \begin{center}
    \subfigure[The warp factor]{
        \includegraphics[width=6cm]{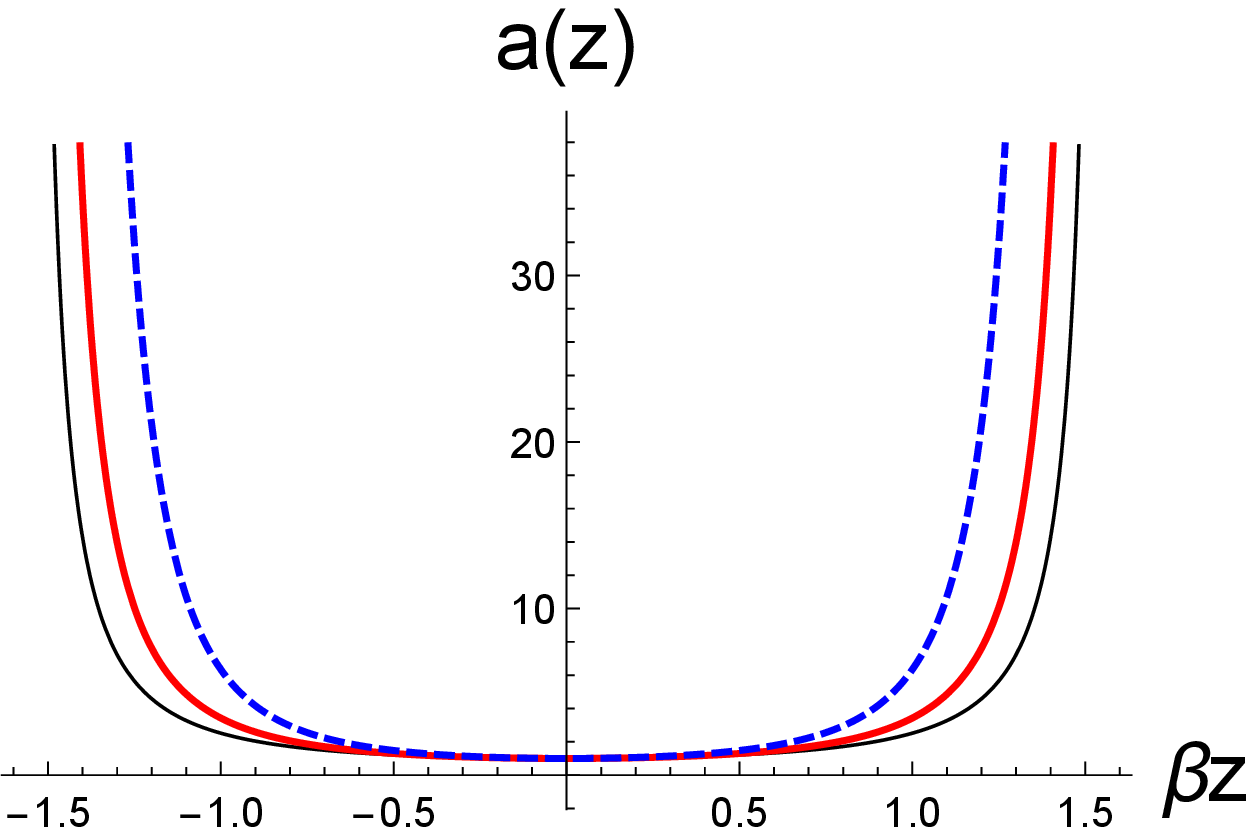}}
    \subfigure[The  scalar field]{
        \includegraphics[width=6cm]{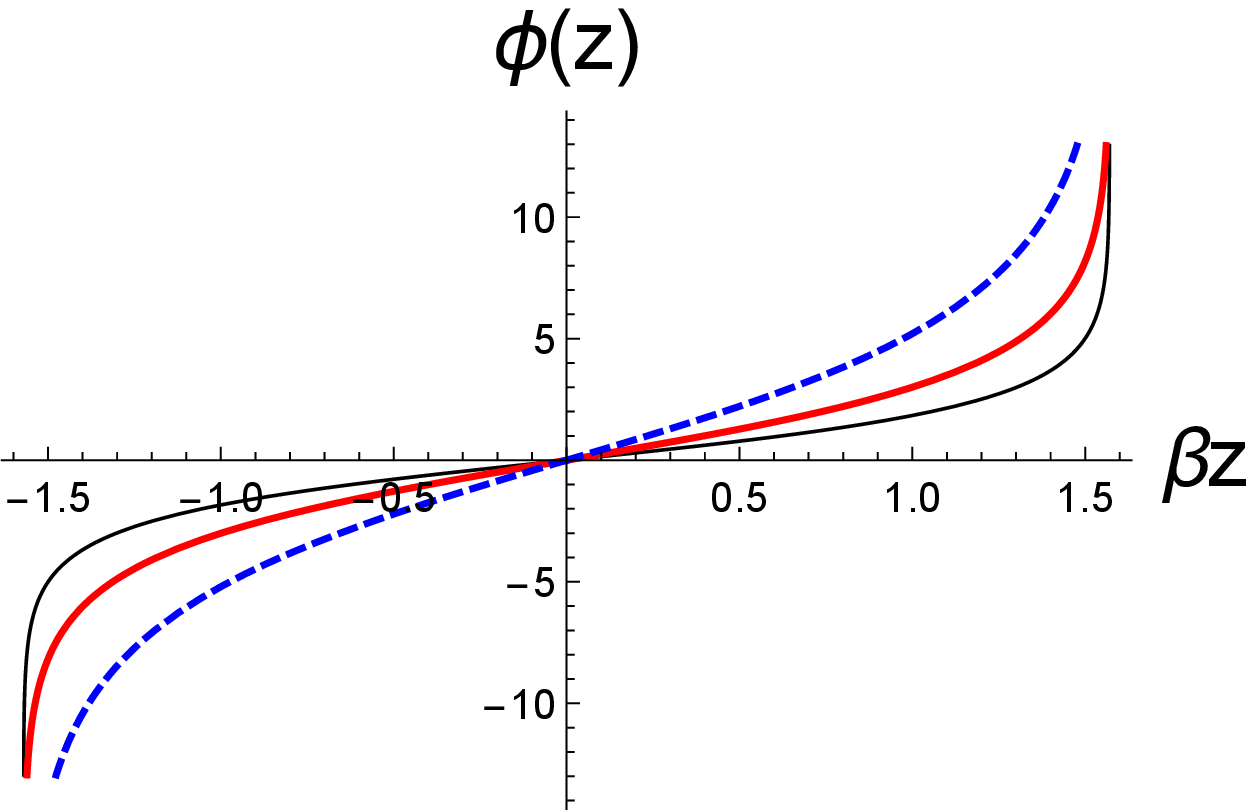}}
    \end{center}
    \caption{The shapes of the warp factor $a(z)$ and the scalar field $\phi(z)$ of the flat brane model. The parameter is set as $p=1.5$ for the thin black lines, $p=2$ for the thick red lines, and $p=3$ for the dashed blue lines.}\label{flat}
    \end{figure}
\section{Tensor perturbation}
In this section, we investigate the tensor perturbation of the Rastall thick brane to analyze the stability and the localization of gravity.
 The perturbed metric of the background is given by
    \begin{eqnarray}
        \tilde{ds}^2=a^2(z) \left[(\gamma_{\mu\nu}+h_{\mu\nu}(x^{\mu},z))dx^{\mu}dx^{\nu}+dz^2\right],
    \end{eqnarray}
where the tensor mode $h_{\mu\nu}$ satisfies the transverse and traceless condition $\eta^{\mu\nu}\partial_{\mu}h_{\alpha\nu}=0$ and $\eta^{\mu\nu}h_{\mu\nu}=0$. The corresponding perturbation of the Ricci tensor $\delta R_{MN}$ is given by
    \begin{eqnarray}
        \label{Ricci1}
        \delta R_{\mu\nu}=&&-\frac{1}{2}\Box^{(4)} h_{\mu\nu}-\frac{1}{2}h_{\mu\nu}''
        -\frac{3}{2}A'h_{\mu\nu}'-(3A'^2+A''-4\epsilon k^2)h_{\mu\nu},
    \end{eqnarray}
where $\Box^{(4)}\equiv\eta^{\mu\nu}\partial_{\mu}\partial_{\nu}$ denotes the four-dimensional d'Alember operator, and the primes denote the derivatives with respect to the extra dimension coordinate $z$.
Substituting Eq. (\ref{Ricci1}) into the perturbation of Eq. (\ref{Rastal eq}), we can obtain the equation of the tensor perturbation $h_{\mu\nu}$,
    \begin{eqnarray}
        \label{tensor}
       -\frac{1}{2}\Box^{(4)} h_{\mu\nu}-\frac{1}{2}h_{\mu\nu}''-\frac{3}{2}A'h_{\mu\nu}'+
       \left[\epsilon k^2+8\lambda(3\epsilon k^2-3A'^2-2A'')\right]h_{\mu\nu}
       =0.
    \end{eqnarray}

 To eliminate the first-order term in Eq. (\ref{tensor}), we define  $h_{\mu\nu}=a^{-\frac{3}{2}}\bar{h}_{\mu\nu}$ and the above equation turns to
    \begin{eqnarray}
        \label{tensor2}
       -\bar{h}_{\mu\nu}''+V_t(z)\bar{h}_{\mu\nu}=(\Box^{(4)}-48 \epsilon\lambda  k^2-2\epsilon k^2) \bar{h}_{\mu\nu},
    \end{eqnarray}
where
    \begin{eqnarray}
        \label{tensor potential}
         V_t(z)=\Big(1-\frac{64}{3}\lambda \Big) \Big (\frac{9}{4}A'^2+\frac{3}{2}A'' \Big).
    \end{eqnarray}
  Applying the Kaluza-Klein decomposition $\bar{h}_{\mu\nu}=\epsilon_{\mu\nu}(x^{\alpha})t(z) $, where the four-dimensional part $\epsilon_{\mu\nu}(x^{\alpha})$ satisfies $(\Box^{(4)}-48 \epsilon\lambda  k^2-2\epsilon k^2) \epsilon_{\mu\nu}(x^{\alpha})= \tilde{m}^2t \epsilon_{\mu\nu}(x^{\alpha})$,  we obtain the Schrodinger-like equation of the tensor mode
    \begin{eqnarray}
        \label{tensor3}
       -t''(z)+V_t(z) t(z)=\tilde{m}^2 t(z),
    \end{eqnarray}
  with the effective potential given by Eq. (\ref{tensor potential}).
  This Schrodinger-like equation of the tensor mode is similar to the one of the thick brane in general relativity, except that the effective potential is modified by a factor $(1-\frac{64}{3}\lambda)$.  When $\lambda=0$, the theory reduces to general relativity and Eq. (\ref{tensor3}) can be  factorized as
      \begin{eqnarray}
        \label{tensor4}
      \Big(\! -\partial_z +\frac{3}{2}A' \Big)\Big(\partial_z +\frac{3}{2}A'\Big)t(z)=m^2 t(z),
    \end{eqnarray}
 of which the solution of the localized mode is
       \begin{eqnarray}
        \label{tensor5}
       t_{0}(z)=\text{e}^{\frac{3}{2}A},
    \end{eqnarray}
and the corresponding mass square is $m_{0}^{2}=0$.

In the Rastall gravity, Eq. (\ref{tensor3}) can not be factorized, which makes it difficult to obtain the solutions of localized modes. Nevertheless, it is possible to deduce the feature of the tensor mode by making use of the knowledge in quantum mechanism.
Comparing Eq. (\ref{tensor3}) with a stationary Schrodinger equation, one can obtain the following Hamiltonian,
    \begin{eqnarray}
       \label{Hamiltonian}
       \tilde{H}=H -\frac{64}{3}\lambda V
    \end{eqnarray}
where $H$ and $V$ are given by
    \begin{eqnarray}
       \label{Hamiltonian}
       H&=&-\frac{d^2}{dz^2}+V,\\
       V&=&\frac{9}{4}A'^2+\frac{3}{2}A''.
    \end{eqnarray}
 The Hellmann-Feynman theorem states
    \begin{eqnarray}
       \label{HF}
       \frac{\partial \tilde{m}_n^2}{\partial \lambda} &=& \frac{\langle t_n | \frac{\partial \tilde{H}}{\partial \lambda} |t_n \rangle}{\langle t_n | t_n \rangle}   \nonumber\\
       &=& -\frac{64}{3}\frac{\langle t_n | V |t_n \rangle}{\langle t_n | t_n \rangle} \nonumber\\
       &\equiv&  -\frac{64}{3}\overline{V}_{n}
    \end{eqnarray}
for an arbitrary bound state $|t_n \rangle$. Since the Rastall parameter satisfies $|\lambda|<<1$, the mass spectrum of the tensor modes is similar with the ones in the general relativity, except for a shift, i.e.
    \begin{eqnarray}
       \label{MS}
       \tilde{m}_n^2\approx m_n^2 -\frac{64}{3}\lambda \overline{V_n}.
    \end{eqnarray}
Note that the prerequisite for the above conclusion is that the bound states of Eq. (\ref{tensor3}) exist.

 Next, we analyze the tensor perturbation further in the cases of flat, de Sitter and anti-de Sitter branes, respectively.

\subsection{Case I: Flat Brane}
\begin{figure}
\begin{center}
\includegraphics[width=8cm]{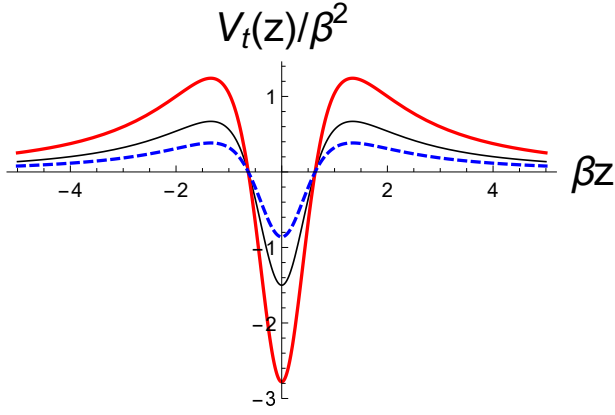}
\end{center}
\caption{ The shapes of the effective potential $V_t (z)$ of the flat brane model. The parameter is set as $\lambda=0$ for the thin black line, $\lambda=-0.04$ for the thick red line, and $\lambda=0.02$ for the dashed blue line. The other parameter is set as $\beta=1.$} \label{vt1}
\end{figure}

In the flat brane model, the warp factor is given by Eq. (\ref{wf1}), and the corresponding effective potential $V_{t}(z)$ reads
    \begin{eqnarray}
       \label{Vt1}
       V_{t}(z)=-\frac{\beta ^2 (64 \lambda -3) \left(5 \beta ^2 z^2-2\right)}{4 \left(\beta ^2 z^2+1\right)^2},
    \end{eqnarray}
which is plotted in Fig. \ref{vt1}. When $\lambda=0$, Eq. (\ref{tensor3}) has and only has one bound tensor mode with eigenvalue $m^2=0$.  Thus, Eq. (\ref{MS}) implies that for $|\lambda| \ll 1$, the mass square of the tensor zero mode is
    \begin{eqnarray}
       \label{MS}
       \tilde{m}_0^2 \approx -\frac{64}{3}\lambda \overline{V_0},
    \end{eqnarray}
where $\overline{V_0}$ is the potential expectation value of the zero mode for $\lambda=0$. Since the corresponding kinetic energy expectation value is positive, we have  $\overline{V_0}<0$.
To prevent the ghost instability, the mass square must satisfy $\tilde{m}_0^2\geq0$, which implies that $\lambda \geq 0$.

On the other hand, it can be seen that $V_t(\infty)=0$, therefore, the mass square $\tilde{m}_0^2$ of the localized mode satisfies $m_0^2\leq 0$. Accordingly for a viable flat brane model, we must demand $\tilde{m}_0^2=0$, which is not possible for nonvanishing $\lambda$. In conclusion, in the flat brane of Rastall gravity, the constraints from the stability and localization yield $\lambda=0$, which just reduce to the case of general relativity.

\subsection{Case II: de Sitter Brane}
In the de Sitter brane model, with the warp factor (\ref{wf2}), the effective potential (\ref{tensor potential}) turns to
   \begin{eqnarray}
       \label{vt ds}
       V_t (z)=\frac{3}{8}p\beta^2 \Big(1-\frac{64}{3}\lambda \Big)
       \left[-(6p+4)\text{sech}^2 (\beta z)+6p \right]
    \end{eqnarray}
Since it is difficult to obtain the analytical solution of  Eq. (\ref{tensor3}) with the above potential, we consider the case that $|\lambda|\ll 1$ and solve this equation by perturbation method.

When $\lambda=0$, it is shown in \cite{PhysRevD.66.024024} that the only localized mode is the zero mode, which can be solved by using the factorizing method,
    \begin{eqnarray}
    	\label{ds zero}
 	    t(z) = \text{sech}^{\frac{3p}{2}}(\beta z).
    \end{eqnarray}

\begin{figure}
\begin{center}
\includegraphics[width=8cm]{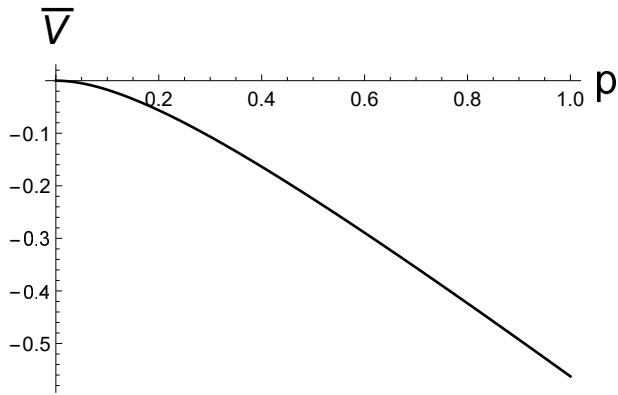}
\end{center}
\caption{ The potential expectation $\overline{V_0}$ as a function of $p$ for the de Sitter brane model. The parameter is set as $\beta=1.$} \label{vp1}
\end{figure}

Then we turn to the case of $\lambda\neq0$. Since $|\lambda| \ll 1$, the  mass square of the bound state is
    \begin{eqnarray}
    \tilde{m}_0^2 \approx -\frac{64}{3}\lambda \overline{V_0},
        \end{eqnarray}
with $\overline{V_0}$ given by
   \begin{eqnarray}
       \label{mass}
       \overline{V_0} &=& \frac{\left\langle t_0 \right| \tilde{H} \left| t_0 \right\rangle}{\left\langle t_0 | t_0 \right\rangle}	\nonumber\\
       		&=& \frac{\int^{\infty}_{-\infty}  t_{0}(z)^2 V(z) dz}{\int^{\infty}_{-\infty}  t_{0}(z)^2 dz}	\nonumber\\
       		&=& \frac{3 \beta ^2 p \Gamma \left(\frac{3 p}{2}+\frac{1}{2}\right) \left(-4 \, _2F_1\left(1,-\frac{3 p}{2}-1;\frac{3 p}{2}+1;-1\right)+\frac{3 \sqrt{\pi } p \Gamma \left(\frac{3 p}{2}+1\right)}{\Gamma \left(\frac{3 (p+1)}{2}\right)}+4\right)}{8 \sqrt{\pi } \Gamma \left(\frac{3 p}{2}\right)},
    \end{eqnarray}
which is plotted in Fig. \ref{vp1}. Using the perturbation method, the corresponding solution of the tensor mode $\tilde{t}(z)$ can be solved, which  is still given by (\ref{ds zero}). It can been seen that $\overline{V_0}<0$, therefore the Rastall parameter must satisfies $\lambda \geq 0$ to ensure that $\tilde{m}_0^2\geq0$.

\begin{figure}
\begin{center}
\includegraphics[width=8cm]{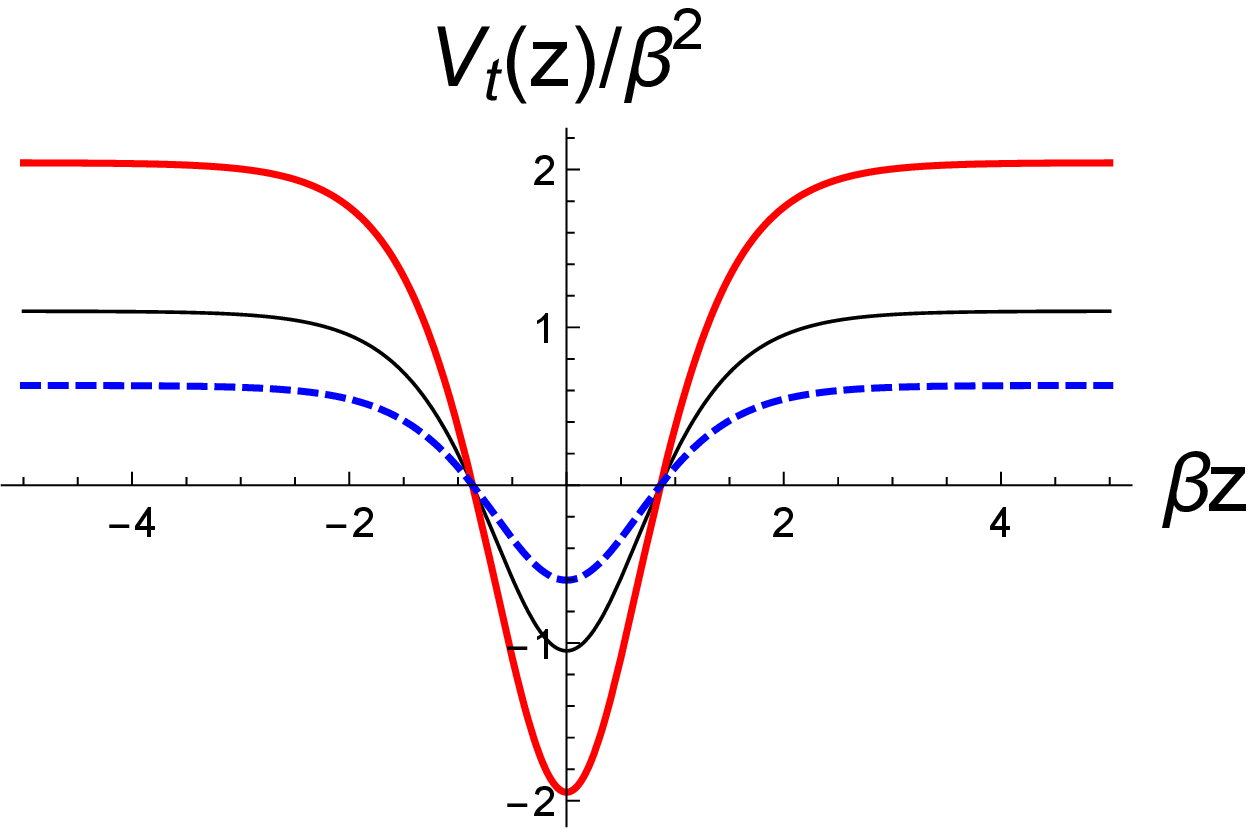}
\end{center}
\caption{ The shapes of the effective potential $V_t (z)$ of the de Sitter brane model. The parameter is set as $\lambda=0$ for the thin black line, $\lambda=-0.04$ for the thick red line, and $\lambda=0.02$ for the dashed blue line.} \label{vt2}
\end{figure}

\subsection{Case III: Anti-de Sitter Brane}
\begin{figure}
\begin{center}
\includegraphics[width=8cm]{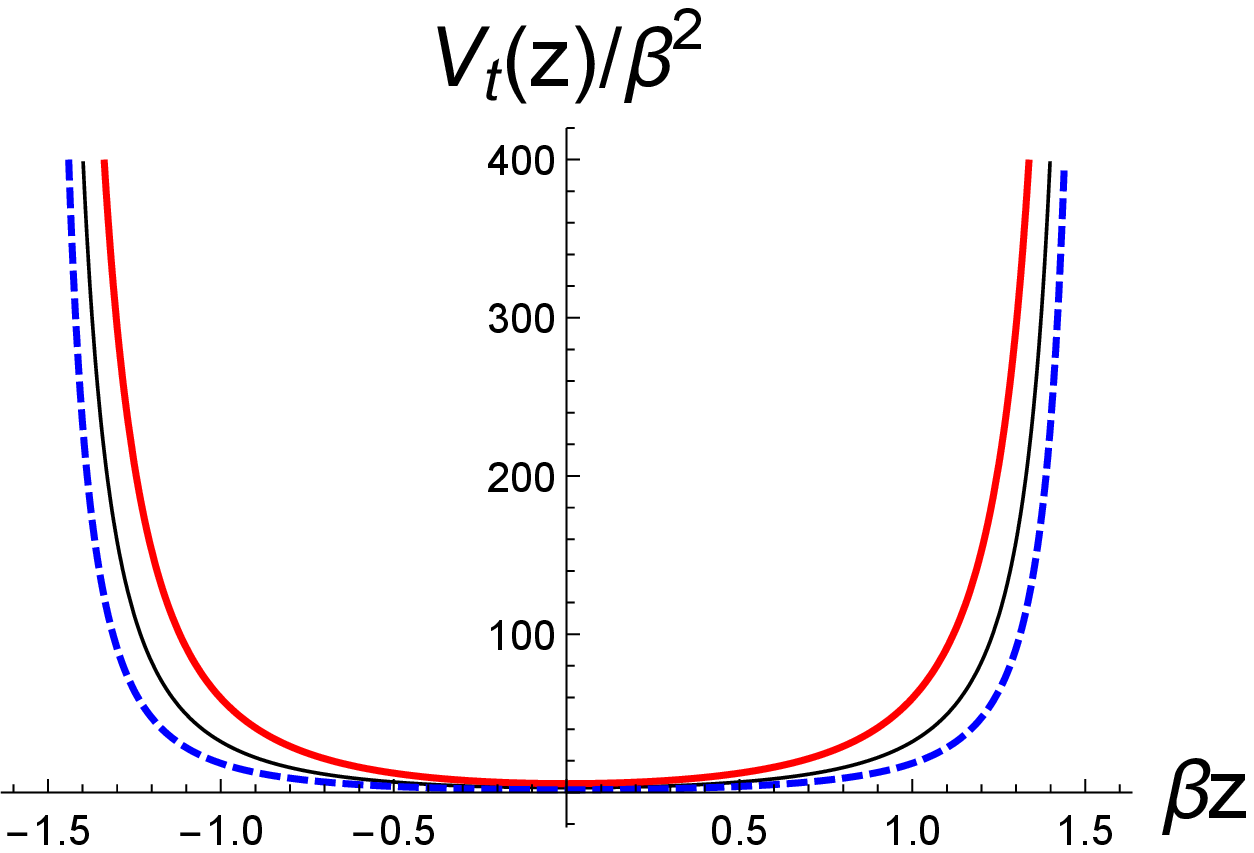}
\end{center}
\caption{ The shapes of the effective potential $V_t (z)$ of the  anti-de Sitter brane model. The parameter is set as $\lambda=0$ for the thin black line, $\lambda=-0.04$ for the thick red line, and $\lambda=0.02$ for the dashed blue line. } \label{vt3}
\end{figure}

In the anti-de Sitter brane model, with the warp factor (\ref{wf3}), the effective potential (\ref{tensor potential}) reads
   \begin{eqnarray}
       \label{vt Ads}
       V_t (z)=\Big(1-\frac{64}{3}\lambda\Big) \frac{3 \beta^2}{4 p}\left[(3 p+2) \sec ^2(\beta  z)-3 p\right],
    \end{eqnarray}
which is plotted in Fig. \ref{vt3}. Since $|\lambda| \ll 1$, $(1-\frac{64}{3}\lambda)>0$, thus the parameter $p$ must satisfy $p<\frac{3}{2}$ to ensure the shape of the potential $V_t (z)$ open upwards.
Again, we solve the tensor perturbation for the case of $\lambda=0$ and obtained the
solution \cite{PhysRevD.84.044033}
   \begin{eqnarray}
       \label{tn Ads}
       t_n (z)=_2\!\!F_1 \left[1-n,1+n+3p,\frac{3(1+p)}{2},\frac{1-\sin(\beta z)}{2}\right]\cos^{1+\frac{3p}{2}}(\beta z),
    \end{eqnarray}
where $n=1,2,\cdots$, and $_2 F_1(c_1,c_2,c_3,z)$ is the hypergeometric function.
And the corresponding mass spetrum is \cite{PhysRevD.84.044033}
   \begin{eqnarray}
       \label{mn Ads}
       m_n^2 =\beta^2 n(n+3p).
    \end{eqnarray}

Therefore, the mass spetrum for the case of $\lambda \ll 1$ is
   \begin{eqnarray}
       \label{mn Ads}
       \tilde{m}_n^2 \approx \beta^2 n(n+3p)-\frac{64}{3}\lambda \overline{V_n}
    \end{eqnarray}
with $\overline{V_n}$ given by
  \begin{eqnarray}
       \label{Vn ads}
       \overline{V_n} = \frac{\int^{z_b}_{-z_b}  t_{n}(z)^2 V(z) dz}{\int^{z_b}_{-z_b}  t_{n}(z)^2 dz}.	
    \end{eqnarray}
The corresponding KK mode can by obtained by using the perturbation method
  \begin{eqnarray}
       \label{tn Ads}
       \tilde{t}_n (z)\approx t_n(z)+\sum_{i} \frac{-\frac{64}{3}\lambda \int^{z_b}_{-z_b}  t_{n}(z) V(z)t_{i}(z) dz}{m_n-m_i}t_i(z).
    \end{eqnarray}
For the ground state of the tensor perturbation, Eq. (\ref{Vn ads}) reads $\overline{V_1}=\frac{9 p^2}{12 p+4}$. Thus the mass $m_1$ reads
   \begin{eqnarray}
       \label{mn Ads}
       \tilde{m}_1^2 \approx \beta ^2 \left[\left(3p+1\right)-\lambda \frac{3 p (9 p+4)}{12 p+4}\right] ,
    \end{eqnarray}
and the solution of the ground state is
   \begin{eqnarray}
       \label{tn Ads}
       \tilde{t}_1 (z) \approx \sqrt{\beta}\pi ^{-\frac{1}{4}} \cos ^{\frac{3 p}{2}+1}(\beta z) \left[c_1-\lambda c_2  \left((3 p+4) \sin ^2(\beta z)-1\right)\right],
    \end{eqnarray}
where
    \begin{eqnarray}
        c_1 &=& \sqrt{\frac{\Gamma \left(\frac{3 p}{2}+2\right)}{\Gamma \left(\frac{3 (p+1)}{2}\right)}} ,\\
        c_2&=&\frac{16 \sqrt{2} p (3 p+2) \sqrt{\frac{(p+2) \Gamma \left(\frac{3 p}{2}+4\right)}{\left(3 p^2+7 p+4\right) \Gamma \left(\frac{3 p}{2}+\frac{5}{2}\right)}}}{(3 p+1) \left(\sqrt{3 p+1}-3 \sqrt{3} (p+1)\right)}.
    \end{eqnarray}
Since the ground state satisfies $m_1^2\geq 0$, the Rastall parameter must satisfy $\lambda<\frac{4(3p+1)^2}{3p(9p+4)}$.

\section{Conclusions and Discussions}
In this paper, we investigated the thick brane model in the Rastall gravity.
In this theory, the thick brane is generated by a nonconserved scalar field, of which the covariant divergence is proportional to the scalar curvature.
 The flat, de Sitter, and anti-de Sitter brane solutions were obtained and the tensor perturbation of the metric was analyzed for each model. We showed that the tensor perturbation in this theory is quite different from that in the general relativity and $f(R,T)$ gravity \cite{Gu:2016nyo}, though the Rastall gravity is equivalent to the $f(R,T)$ gravity with the Lagrangian given by $f(R,T)= R+\alpha T$ \cite{Shabani:2020wja}.
The new features of the tensor perturbation in the Rastall thick brane models come from that the effective potential of the Schrodinger-like equation is multiplied by the factor $1-\frac{64}{3}\lambda$.

In the flat brane model, for the case $\lambda\neq 0$, it was proved that the tensor mode of the perturbation is either unstable or nonlocalizable. In the de Sitter brane model, only the ground state of the tensor mode is localized, and the mass and solution of the ground state were obtained by the perturbation method. In the anti-de Sitter brane model, the number of the bound state is infinity. For both the de Sitter and anti-de Sitter brane models, the condition of stability for the parameter $\lambda$ was obtained.


\section*{Acknowledgement}

Y. Zhong acknowledges the support of the Fundamental Research Funds for the Central Universities (Grants No.531118010195) and the National Natural Science Foundation of China (Grant No. 12047501), he also thanks the generous hospitality offered by the Center of Theoretical Physics at Lanzhou University where part of this work was completed.  K. Yang acknowledges the support of the National Natural Science Foundation of China under Grant No. 12005174, and Natural Science Foundation of Chongqing, China under Grant No. cstc2020jcyj-msxmX0370.

\bibliographystyle{JHEP}


\providecommand{\href}[2]{#2}\begingroup\raggedright\endgroup

\end{document}